# Exoplanet Magnetic Fields

**David A. Brain**
*Laboratory for Atmospheric and Space Physics*
*University of Colorado Boulder*
*Boulder, CO 80303, U.S.A.*
*david.brain@colorado.edu*

**Melodie M. Kao**
*Department of Astronomy and Astrophysics*
*University of California Santa Cruz*
*Santa Cruz, CA 95064*
*melodie.kao@ucsc.edu*

**Joseph G. O'Rourke**
*School of Earth and Space Exploration*
*Arizona State University*
*Tempe, Arizona 85287, U.S.A.*
*jgorourk@asu.edu*

## INTRODUCTION

Planetary magnetic fields are important indicators of planetary processes and evolution, from a planet's outer core to its surface (if it possesses one) to its atmosphere and near-space environment. Magnetic fields are most directly measured in situ, and determining whether distant planetary objects possess magnetic fields can be challenging. At present we have no unambiguous measurements of magnetic fields on exoplanets. Nevertheless, it would be surprising if at least some exoplanets did not generate a magnetic field, like many planetary bodies in the solar system.

This chapter provides an overview of the current understanding of exoplanetary magnetic fields and their consequences. In the next section we review the current understanding of planetary dynamo generation as it applies to solar system objects and discuss the implications for exoplanetary magnetic field generation. Following this, we describe seven methods for determining the existence and strength of an exoplanetary magnetic field and discuss the near-term prospects for each method. We close by highlighting four main consequences of exoplanetary magnetic fields for a planet and its evolution.

## THEORY OF EXOPLANET MAGNETIC DYNAMOS

A dynamo turns fluid motion into a magnetic field via electromagnetic induction. In (exo)planetary science, theory aims to answer three questions arising from this definition. First, what planetary bodies host large reservoirs of electrically conductive fluid? The discussions below will largely focus on dynamos in terrestrial planets, but the basic principles apply to gas giants as well. Second, what processes can provoke this fluid into motion? In Earth, our long-lived dynamo arises from thermo-chemical convection in the deep interior (e.g., Landeau et al., 2022). Third, what is the strength and geometry of the resulting magnetic fields? Models of exoplanets often assume that the dipole moment of a dynamo-generated field is aligned with the rotational axis of the planet. On average over geologic timescales, Earth indeed has a geocentric axial dipole (e.g., Meert, 2009). However, short-term variations in orientation are common (~10° for Earth now) – and the present-day dynamos of Neptune and Uranus are severely misaligned with their rotational axes today.

Ultimately, answering questions about planetary dynamos requires a diversity of disciplines—not only the physics of electromagnetism, but also mineral physics, geochemistry, and geodynamics. Progress arises from theory, numerical simulations, and laboratory experiments. Beyond the short introduction here, readers are also encouraged to consult more comprehensive reviews of the theory of (exo)planetary dynamos (e.g., Stevenson, 2003, 2010; Christensen, 2010; Schubert & Soderlund, 2011; Laneuville et al., 2020).

**Fundamental Dynamo Equations**

Mathematically, dynamo theory connects the fluid velocity field (**v**) to the magnetic field (**B**). Three variables describe the thermodynamic properties of the fluid: density ($\rho$), pressure ($P$), and temperature ($T$). At least five equations with vectors (more for scalar quantities) are thus needed to solve for these variables. First, we can consider the induction equation, which combines Maxwell's equations with the Lorentz force law and Ohm's law (e.g., Schubert & Soderlund, 2011):

$$\frac{\partial \mathbf{B}}{\partial t} = \nabla \times (\mathbf{v} \times \mathbf{B}) - \nabla \times (\eta \nabla \times \mathbf{B}). \qquad (1)$$

Here, the magnetic diffusivity is $\eta = 1/(\mu_0 \sigma)$, where $\mu_0 = 4\pi \times 10^{-7}$ H/m and $\sigma$ is the electrical conductivity (in units of S/m). The first and second terms on the left side, respectively, relate to the convection and diffusion of the magnetic field.

Next, we can express the conservation of mass as

$$\frac{\partial P}{\partial t} + \nabla \cdot (\rho \mathbf{v}) = 0, \qquad (2)$$

meaning that mass can move around but not be spontaneously created or destroyed. Then, the conservation of momentum is

$$\rho \frac{D\mathbf{v}}{Dt} + 2\rho \mathbf{\Omega} \times \mathbf{v} = -\nabla P - \rho \mathbf{\Omega} \times (\mathbf{\Omega} \times \mathbf{r}) + \frac{1}{\mu_0}(\nabla \times \mathbf{B}) \times \mathbf{B} + \rho g + \frac{1}{3}\rho \nu \nabla(\nabla \cdot \mathbf{v}) + \rho \nu \nabla^2 \mathbf{v}, \qquad (3)$$

where the rotation rate of the reference frame is $\mathbf{\Omega}$; the position vector is $\mathbf{r}$; the gravitational acceleration is $g$; and the kinematic viscosity is $\nu$ (Schubert & Soderlund, 2011). From left to right, there are eight terms in Eq. 3 (Schubert & Soderlund, 2011): inertial acceleration, Coriolis acceleration, pressure gradient, centrifugal force, Lorentz force, buoyancy force, and two terms related to viscous diffusion under the so-called Stokes assumption (i.e., neglecting the bulk viscosity).

Fourth, we can express the conservation of energy as

$$\rho C_P \frac{DT}{Dt} - \alpha T \left(\frac{DP}{Dt}\right) = \Phi + \nabla \cdot (k \nabla T), \qquad (4)$$

where time is $t$; the coefficient of thermal expansion is $\alpha$; the specific heat at constant pressure is $C_p$; the total dissipation (viscous and ohmic heating) is $\Phi$; and the thermal conductivity is $k$ (Schubert & Soderlund, 2011). The right side of this equation is the dissipation associated with the dynamo. The left side is the change in thermal energy, including the effects of adiabatic (de)compression.

Last, we need an equation of state:

$$\rho = f(P, T), \qquad (5)$$

where $f$ is whatever function is typically used for the material in question (e.g., metallic hydrogen, molten silicates, iron alloys, etc.). Solving these equations is difficult, even with modern numerical approaches. To train our intuition for planetary dynamos, we should first analyze non-dimensional parameters that arise from these fundamental equations.

**Key Non-Dimensional Parameters**

Non-dimensional parameters are ratios between different forces or other quantities. They reveal the terms that are the most important in the governing equations. Numerical models that solve Eqs. 1–5 need five control parameters for a dynamo. Two other parameters provide criteria for the existence (or lack thereof) of dynamos in different types of planets (e.g., Stevenson, 2003, 2010).

***Control Parameters for Numerical Models of Dynamos.*** Control parameters help characterize numerical models of dynamos. For example, we can define five non-dimensional parameters (the Rayleigh, Reynolds, Ekman, Prandtl, and magnetic Prandtl numbers) that are popular among modelers (e.g., Schubert & Soderlund, 2011):

$$Ra = \frac{\alpha g \Delta T D^3}{\nu \kappa}; \quad Re = \frac{vD}{\nu}; \quad E = \frac{\nu}{\Omega D^2}; \quad Pr = \frac{\nu}{\kappa}; \quad Pr_m = \frac{\nu}{\eta}. \tag{6}$$

The Rayleigh number ($Ra$) is the ratio of the timescales for the transport of heat via diffusion and convection. High $Ra$ (e.g., $> 10^3$) indicates vigorous convection. Here, $\Delta T$ is a temperature difference that drives convection; $D$ is a length scale; and $\kappa$ is the thermal diffusivity of the fluid. The Reynolds number ($Re$) is the ratio of inertial and viscous forces in the fluid. Dynamo-generating flows tend to be turbulent with $Re > \sim 10^3$–$10^4$, typically. In contrast, the Ekman number ($E$) is the ratio of viscous and Coriolis forces in the fluid. The Prandtl number ($Pr$) is the ratio of the diffusivity of momentum (kinematic viscosity) and of heat, while the magnetic Prandtl number ($Pr_m$) is the kinematic viscosity divided by the magnetic diffusivity. While these parameters control the properties of an extant dynamo, they do not quantify the prerequisites for a dynamo to exist.

***Criteria for the Existence and Strength of a Planetary Dynamo.*** A dynamo requires a continual source of power, such as the flux of heat from (hot) planets to (cold) space. The five non-dimensional parameters in Eq. 6 control the properties of an active dynamo. To determine if a dynamo should exist at all, we need to define two other parameters (e.g., Stevenson, 2003, 2010):

$$Re_m = \frac{vD}{\eta}; \quad Ro = \frac{v}{\Omega D}. \tag{7}$$

Here, the magnetic Reynolds number ($Re_m$) is the ratio of magnetic induction and diffusion. The Rossby number (Ro) is the ratio of inertial and Coriolis forces. A dynamo requires $Re_m > 10$–$100$ (so the field is strengthened faster than it diffuses away) and Ro $\ll 1$ (so the Coriolis force is dynamically important) (e.g., Stevenson, 2003, 2010). In practice, the Rossby-number criterion is satisfied for even the slowest-rotating planet in the Solar System (Venus) and thus also for tidally locked exoplanets with close-in orbits (such as hot Jupiters). Therefore, the first criterion is most important (e.g., Stevenson, 2003, 2010). A dynamo is expected when a large (big $D$) volume of conductive fluid (small $\eta$) is undergoing convection (non-zero $v$).

Figure 1 shows how $Re_m$ yields a "regime diagram" for dynamos that explains why they are marginal in terrestrial planets but widespread in gas and ice giants. If convection stops, then the dynamo dies on the timescale of magnetic diffusion (e.g., Stevenson, 2003, 2010), i.e., $\tau \sim D^2/(\pi^2 \eta)$ (e.g., $\sim 10^4$ and $\sim 10^8$ years for Earth and Jupiter, respectively). Since that timescale is a geologic eyeblink, observation of a live dynamo basically means that convection is currently occurring. Overall, theory has taken us from the induction equation to the realization that studying planetary dynamos means analyzing when and where convection is likely to happen.

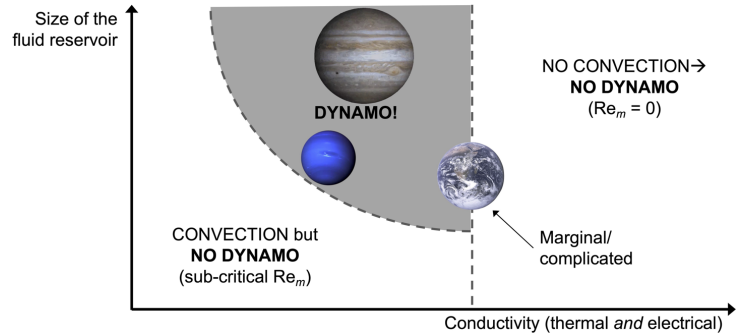

Figure 1: Prospects for a dynamo based on the magnetic Reynolds number. Only in the grey, shaded region is $Re_m$ large enough for the dynamo. Below and to the left, $Re_m$ is non-zero but too small for a dynamo. The fluid is not moving fast enough and/or the reservoir is too small. The absence of a dynamo in the rightmost region is perhaps counter-intuitive. A fluid is unlikely to host a dynamo if it is too electrically conductive (e.g., if Earth's core were made of copper or gold) because it would be very thermally conductive as well—and thus likely to transport heat via conduction without fluid motions. Therefore, gas and ice giant planets plot in more favorable regions, even though their interiors are less electrically conductive than Earth's metallic core. Terrestrial planets are marginal in their capacity to host a dynamo (e.g., Earth plots at the same location as dynamo-less Venus). Created based on Figure 1 in Stevenson (2003).

## Application of Theory to Earth's Dynamo

We can estimate the control parameters for perhaps the best-studied planetary dynamo: the one that exists today in Earth's outer core. For liquid iron alloys at the relevant pressures and temperatures, $v \sim 10^{-6}$ m$^2$/s, $\eta \sim 1$ m$^2$/s, and $\kappa \sim 10^{-5}$ m$^2$/s (e.g., Schubert & Soderlund, 2011). Therefore, $Pr_m \sim 10^{-6}$ and $Pr \sim 0.1$ for Earth's outer core. Because $Pr \gg Pr_m$, magnetic fields vary over much longer length scales than variations in the velocity field. Earth's 24-hour day means $\Omega \sim 7.3 \times 10^{-5}$ s$^{-1}$, so $E \sim 10^{-15}$ for our core with $D \sim 3000$ km. Flow velocities in Earth's outer core are inferred to be $v \sim 0.5$ mm/s (e.g., Bloxham & Jackson, 1991), meaning that $Re > 10^7$. As discussed below, $\Delta T$ for the core is unknown—and perhaps the relevant buoyancy is chemical, not thermal. However, estimates for Ra have ranged from $\sim 10^{22}$–$10^{30}$ (e.g., Schubert & Soderlund, 2011). Ultimately, convection in Earth's outer core is vigorous and turbulent—and the Coriolis force governs the patterns of fluid motion.

Numerical simulations would require impossibly small spatial and temporal resolution to implement realistic control parameters. For example, a cutting-edge study of Earth's dynamo used $Ra \sim 5 \times 10^6$; $E = 3 \times 10^{-4}$; $Pr = 1$; and $Pr_m = 10$ (Driscoll, 2016). Those assumed control parameters were at least $\sim$1–15 orders of magnitude away from the best estimates of their true values.

These numerical studies are still worthwhile! Published dynamo models are "Earth-like" in many respects, especially regarding the dipolarity of the magnetic field and its strength at the surface (e.g., Kuang & Bloxham, 1997; Christensen et al., 2010). People can extrapolate scaling laws for dynamo properties, using scaling laws derived by running dynamo models over the accessible ranges of control parameters. However, we cannot use even "Earth-like" dynamo models to simulate the billion-year lives of planetary dynamos. Resource limitations mean that they can model no more than a few million years at a time (e.g., Matsui et al., 2016). As shown in Figure 2, scientists use three-dimensional models of dynamos to provide "snapshots" of their behavior (e.g., Driscoll, 2016). We must turn to the simpler criteria to study how Earth's dynamo changed over geologic time as the core evolved.

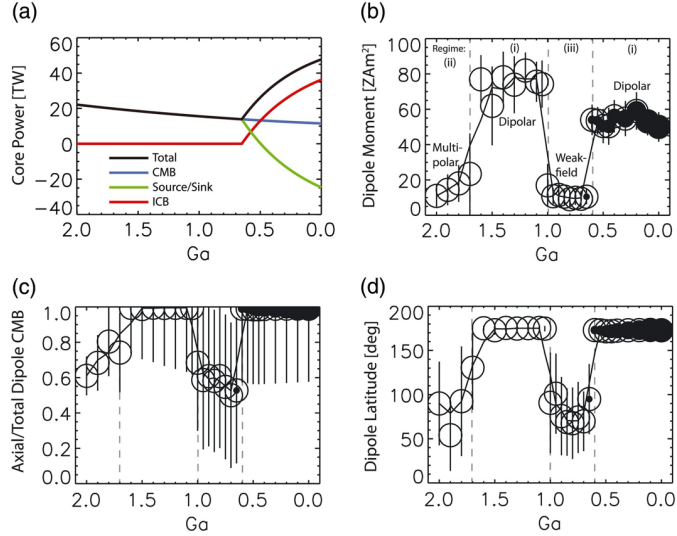

Figure 2: Combining thermal evolution models with numerical simulations of dynamos. (a) One-dimensional, parameterized models calculate how fast the core cools over time; the size of the inner core; and the magnitude of the dissipation (power) available for a dynamo over billions of years. Dynamo models then predict the strength (b), dipolarity (c), and tilt (d) of the magnetic field at representative times. This is Figure 2 from Driscoll (2016), reproduced with permission.

We know that Earth's dynamo exists today because we are immersed in the resulting magnetic field. Using the parameters quoted above, $Re_m \sim 1500$ and $Ro \sim 2 \times 10^{-6}$, which easily satisfies the criteria from the subsection above. The magnetic diffusion time for Earth's dynamo is $\sim 10^4$ years (e.g., Stevenson, 2003, 2010), while we know from paleomagnetism that the dynamo has existed for at least $\sim 3.5$ billion years (e.g., Tarduno et al., 2010) – and possibly since soon after Earth accreted (c.f., Borlina et al., 2020). However, the difficulty in understanding Earth's dynamo lies in explaining why fluid velocities in the liquid outer core have been non-zero.

Earth's dynamo can live if convection remains the dominant mode of heat transfer in the fluid. In planetary interiors, the first enemy of convection is thermal conduction. The mode that "wins" is the one that can transport heat most rapidly in a given situation. Thus, a dynamo will die if the cooling rate of the fluid is low enough that conduction can transport the associated heat flux. Quantitatively, convection in a low-viscosity fluid will tend to maintain an adiabatic thermal gradient, i.e., $(dT/dr)_{ad} \sim -\alpha T g/C_p$, where $r$ is distance from the planetary center (e.g., Stevenson, 2003, 2010). By Fourier's law, the associated "adiabatic heat flux" is $Q_{ad} = -k(dT/dr)_{ad}$, where $k$ is the thermal conductivity of the fluid. In thermal evolution models, a simple way to assess the likelihood of a dynamo is to compare the actual heat flux out of the fluid region ($Q$) to the adiabatic heat flux. If $Q > Q_{ad}$, then thermal convection can sustain a dynamo, assuming the fluid is homogeneous and not undergoing any chemical reactions. Otherwise, convection will cease as heat moves towards the top of the fluid and creates thermal stratification. Although this criterion is easy to impose in models, we do not know if it holds for Earth's core at present day.

Earth's outer core is possibly cooling too slowly today for thermal convection alone to sustain our dynamo. The adiabatic thermal gradient in the core is roughly $(dT/dr)_{ad} \sim 0.5$ K/km (e.g., Stevenson, 2003, 2010; Schubert & Soderlund, 2011). The thermal conductivity of Earth's core remains debated, with published estimates ranging from $\sim 30$–$200$ W/(m K) (e.g., Williams, 2018). Thus, the adiabatic heat flux might be $\sim 15$–$100$ mW/m². In comparison, the vigor of convection in the overlying, silicate mantle determines $Q$ for the core. Estimates span $Q \sim 30$–$100$ mW/m² (e.g., Lay et al., 2008). Obviously, these estimates of $Q$ and $Q_{ad}$ overlap. If higher estimates of $k$ are proved correct, then thermal convection cannot power Earth's dynamo without assistance.

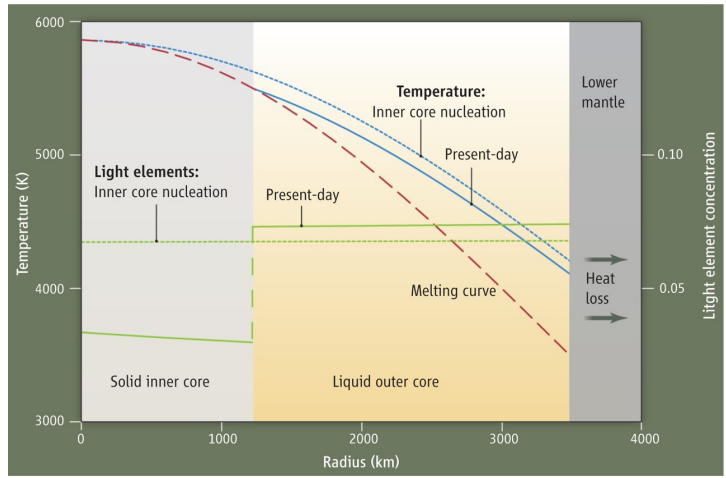

Figure 3: Earth's center is the hottest place in the planet, but it is slowly freezing because of the extreme pressure. Growth of the inner core (left, grey region) excludes light elements into the outer core (green lines). The inner core boundary is the intersection between the adiabatic temperature profile (blue curves) and the melting temperature (red curve). Earth has an inner core because the melting curve and the adiabat first cross at the center. The fluxes of heat and light elements both provide power for the dynamo. If the total heat flux is sub-adiabatic at present day, a thermally stratified layer could exist at the top of the core (not shown here). This is the Figure from Olson (2013), reproduced with permission.

Fortunately, we need not panic that Earth's dynamo might die soon. The inner core is our dynamo's savior because it provides chemical buoyancy to power convection (Figure 3). As the core cools, the inner core slowly freezes and grows, excluding "light elements" from its solid structure and injecting them into the bottom of the outer core. This compositional buoyancy is enough to drive convection (e.g., Labrosse, 2015). Even if $Q < Q_{ad}$, chemical convection can move hot fluid downwards and cold fluid upwards. Scaling laws show that the total energy flux—any heat flux from thermal convection plus the gravitational energy associated with chemical convection—determines the strength of the dynamo (e.g., Christensen, 2010). However, the inner core is almost certainly much younger than Earth's dynamo (e.g., Labrosse, 2015; Landeau et al., 2017, 2022). Therefore, explaining the operation of the dynamo in the past is more difficult than accepting its present-day existence.

Scientists coined the term "the new core paradox" to encapsulate the seeming disconnect between the inner core's relative youth and its (perhaps) singular role in maintaining the dynamo (Olson, 2013). Several solutions are under active investigation (e.g., Landeau et al., 2022; Driscoll & Davies, 2023). Briefly, other chemical processes (such as the precipitation of light species such as MgO, $SiO_2$, and/or FeO) could provide chemical buoyancy in the core prior to the nucleation of the inner core (e.g., O'Rourke & Stevenson, 2016; Badro et al., 2016; Hirose et al., 2017). Additionally, the metallic core is perhaps not the only place that could have hosted a dynamo. Earth's silicate mantle might have solidified from the middle outwards after accretion (e.g., Labrosse et al., 2007; Coltice et al., 2011; Pachhai et al., 2022; Ferrick & Korenaga, 2023), rather than from the bottom upwards as traditionally assumed. A "basal magma ocean" could have survived for roughly half of Earth's life at the base of the mantle (e.g., Labrosse et al., 2007). Scientists recently learned that molten silicates are electrically conductive under deep mantle pressures and temperatures (e.g., Stixrude et al., 2020). If the basal magma ocean was vigorously convecting, then it could have hosted a dynamo early in Earth's history (Ziegler & Stegman, 2013; Blanc et al., 2020), when a core-hosted dynamo seems less likely. Recent studies suggested that a basal magma ocean exists today in Mars (Khan et al., 2023; Samuel et al., 2023), in the past in the Moon (Scheinberg et al., 2018; Hamid et al., 2023), and in the past (and perhaps today) in Venus (O'Rourke, 2020). Basal magma oceans are probably even larger and longer-lived in massive, terrestrial exoplanets, known as super-Earths (e.g., Soubiran & Militzer, 2018). There are no true paradoxes, only gaps in our scientific understanding.

Astute readers might realize that a new core paradox implies the existence of an "old core paradox." Indeed, half a century ago, scientists recognized a "paradox" because mineral physics experiments indicated that the core should not first freeze at its center, if the liquid were vigorously convecting (Kennedy & Higgins, 1973). Since the existence of the inner core (from seismology) and the dynamo (from daily experience) were irrefutable, scientists were not surprised to find an error with the experiments (e.g., Olson, 2013). Modern measurements of the adiabatic gradient and the melting curve of iron alloys are now self-consistent (e.g., Anzellini et al., 2013). Still, the lesson remains that our understanding of planetary dynamos can hinge on quantities that are extraordinarily difficult to determine. Depending on the relative slopes of the melting curve and the adiabat in a particular fluid, crystallization can occur at the top, middle, or bottom of a fluid reservoir (Figure 4). Some critical constraints (such as the existence of an inner core) are likely impossible to obtain for exoplanets in the foreseeable future.

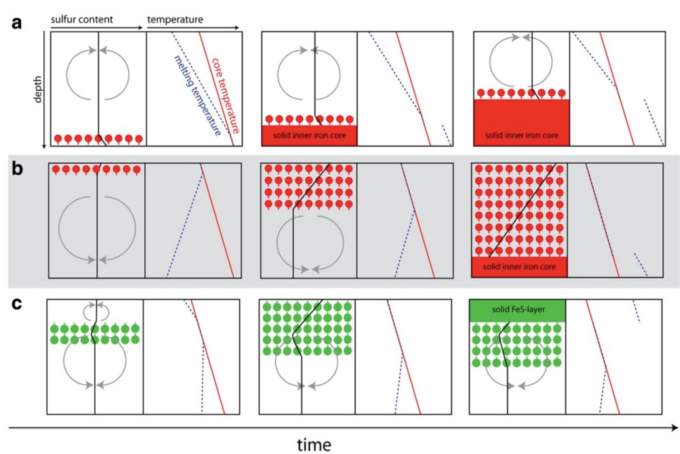

Figure 4: Metallic cores do not always solidify from the center outwards. (a) An Earth-like regime with an inner core is probably the most favorable for a dynamo. (b) Iron snow happens if the adiabat first crosses the liquidus at the top of the core. This regime might be common in smaller terrestrial worlds. (c) In some cores, the crystallizing solid might be less dense than the residual fluid. For example, a core with a high sulfur content, above the Fe-FeS eutectic composition, could crystallize solid FeS that rises to the top. Other regimes are possible. Overall, chemistry is critical to the chances for a dynamo in terrestrial worlds. This is a truncation of Figure 1 from Breuer et al. (2015), reproduced under its open access license.

**Application of Theory to the Solar System**

Many questions await answers about Earth's dynamo, but theory must also attempt to explain a puzzling array of observations from across the Solar System (Table 1). For example, gas and ice giants in the Solar System all have dynamos today. The dynamos in the ice giants are strikingly multipolar and non-axisymmetric (e.g., Stanley & Bloxham, 2004; Soderlund & Stanley, 2020), raising many questions that await answers from new missions (e.g., Cohen et al., 2022). Jupiter's dynamo field is actually better mapped than Earth's dynamo field (e.g., Stevenson, 2010; Moore et al., 2018; Connerney et al., 2022), because Earth's magnetic field includes contributions from magnetized minerals in the crust. This crustal remanent magnetism is how we know that Earth had a dynamo in the past. Obviously, studying remanent magnetism requires a solid piece of a planetary body, either in situ in the crust or as a meteorite. So, we cannot prove that the giant planets had dynamos before human history, but we have no reason to believe that they did not. Based on the reasoning that underlies Figure 1, we expect near-ubiquitous dynamos in giant exoplanets, although their strength, orientation, and dipolarity are uncertain.

Empirically, dynamos in terrestrial planets are fragile things. No planetary property clearly correlates with where dynamos are found in terrestrial bodies in the Solar System—neither size nor bulk composition nor distance from the Sun. It seems almost random that Mercury and Ganymede host dynamos today while Venus and Mars do not. The Solar System also lacks an example of large terrestrial exoplanets (super-Earths), which are apparently common in the galaxy. Super-Earths could be relatively likely to host dynamos simply due to their size (e.g., van Summeren et al., 2013; Boujibar et al., 2020; Blaske & O'Rourke, 2021). However, a higher mass leads to greater pressures in their deep interiors. High pressure could decrease the convective vigor of the solid mantle (by making it more viscous)—and thus slow the cooling rate of any deeper fluid region(s) (e.g., Tachinami et al., 2011). Overall, we should expect lots of dynamo-related diversity even among exoplanets with similar sizes, compositions, and orbits.

**Table 1.** Terrestrial bodies in the Solar System, grouped based on the(non)existence of a dynamo now and/or in the past. Searches for remanent magnetization in meteorites and in the crusts of terrestrial bodies allow us to learn if a planetary body had a dynamo in the past, even if one does not exist today. The existence of metallic cores in many icy satellites, such as Europa, is uncertain.

| Observational Constraints | Planetary Body |
| --- | --- |
| Dynamo exists today | Mercury, Earth, Ganymede, Jupiter, Saturn, Uranus, Neptune |
| Dynamo existed in the past | Mercury, Earth, Moon, Mars, sundry meteorite parent bodies |
| Metallic core, but no known dynamo | Venus, Io, Europa (?) |

We can tell myriad stories about individual worlds. Frustratingly, explanations for the presence or absence of a dynamo often hinge on relatively minor differences. Convection has many enemies. For example, a canonical explanation for Venus's lack of a dynamo is that its slightly lower internal pressures, relative to Earth, fatally delayed the nucleation of the inner core (Stevenson et al., 1983). Or perhaps Venus's core formed with a stabilizing chemical gradient that would inhibit convection even if it were cooling faster than Earth's core (Jacobson et al., 2017). Radiogenic heating in the silicate mantle (from the slow decay of K, U, and Th isotopes) can also kill a dynamo by slowing the cooling rate of the deep interior (e.g., Nimmo et al., 2020). Tidal heating promotes geologic activity near the surface but can also suppress the cooling of the core, which is likely why Io does not have a dynamo today (e.g., Wienbruch & Spohn, 1995). Ultimately, understanding dynamos in terrestrial planets means understanding terrestrial planets as complex systems. The number of potentially pivotal parameters is larger than the number of terrestrial planets in the Solar System, so observational constraints from exoplanets will be invaluable.

**Takeaway Messages for Studies of Exoplanets**

The theory of planetary dynamos offers perhaps only two firm predictions for future studies of exoplanets. First, dynamos are likely quite common, if not universal, in gas and ice giant planets. Measuring the dipolarity and orientation of such dynamos in exoplanets would yield new insights into our understanding of the internal structure and dynamics of these worlds—in parallel with the ongoing exploration of Jupiter, by the NASA Juno mission, and a future mission to an ice giant. Second, some terrestrial planets will host dynamos while others will not. Understanding dynamos in terrestrial planets requires as much geochemistry as electrodynamics, given the critical role of chemical convection. Exoplanets offer a large sample size that will enable statistical tests of which planetary properties most influence the likelihood of a dynamo. Dynamos are thus one of the best examples of how studying exoplanets will teach us about the Solar System, as highlighted in the newest decadal survey, "Origins, Worlds, and Life: A Decadal Strategy for Planetary Science and Astrobiology 2023-2032." Our ignorance about the dynamos of terrestrial planets in particular might be the best motivation for an ambitious observational campaign in the coming decades.

# ASSESSING EXOPLANET MAGNETIC FIELDS

Methods for detecting and assessing exoplanet magnetic fields broadly fall into two categories: direct measurements and indirect inferences of magnetic properties. Table 2 summarizes these methods. In principle, most may apply to terrestrial planets except where noted. In practice, these methods largely favor gas giants for current and/or next generation instruments. The following two subsections give brief overviews on the relevant theory, advantages, limitations, and notable observational efforts. This discussion is by no means a complete review but aimed toward giving the interested reader a general sense of the current measurement landscape and a starting point for a deeper dive into the literature.

Table 2. Summary of exoplanet magnetic field measurement methods.

| Method | Planet type | Information |
|---|---|---|
| **Direct** | | |
| Exoplanet aurorae | all | local strength |
| He 1083 nm spectropolarimetry | transiting hot Jupiter | l.o.s. averaged strength |
| Radiation belt emission | all | dipole magnetic component |
| **Indirect** | | |
| Star-planet interactions | close-in | magnetopause size |
| Ohmic dissipation | transiting hot Jupiter | |
| Magnetospheric bow shocks | transiting | magnetopause size |
| Atmospheric outflow transit spectroscopy | transiting close-in | strongly or weakly magnetized |

**Direct Measurements**

*Exoplanet Aurorae.* The discovery of Jupiter's radio aurorae[1] by Burke & Franklin (1955) precipitated a decades-long search for exoplanet radio aurorae that continues today (Turner et al., 2021, 2023, and references therein). Solar System planets demonstrate several mechanisms for producing aurorae: Most familiar are aurorae on Earth, Jupiter, and Saturn driven by incident plasma flow from the solar wind. An analogous mechanism also occurs on the Galilean moons, where Jupiter's circumplanetary plasma torus supplies the incident plasma (de Kleer et al., 2023). Additionally, magnetospheric plasma departing from rigid co-rotation couples Jupiter's outer magnetosphere with its ionosphere to produce its main aurorae (Cowley & Bunce, 2001; Nichols & Cowley, 2003). Finally, close-in satellites like Io, Europa, Ganymede, Enceladus and perhaps Callisto can excite aurorae on their hosts (Clarke et al., 2002, 2011).

---

[1] The term "aurorae" typically refers to atomic and molecular emissions in a planet's upper atmosphere in response to energy deposition from current systems driven by the described mechanisms. Here, "radio aurorae" refer to radio emissions from electrons in these auroral current systems.

Each of these mechanisms produce multiwavelength aurorae, but it is the radio component of auroral emissions that offers a direct measurement of an object's magnetic field strength. In rarified plasmas where the plasma frequency ($\propto n_e^{1/2}$) is much less than the cyclotron frequency ($\propto B$), upward traveling and mildly relativistic electron populations with energy anisotropies can excite the electron cyclotron maser instability rather than plasma emission (Treumann, 2006). This instability produces periodically repeating and extremely bright, coherent radio emissions easily distinguished from incoherent (gyro)synchrotron emissions via high brightness temperatures and exceptionally strong circular polarization (Dulk, 1985; Hallinan et al., 2008).

When plasma electron number densities $n_e$ are low, radio aurorae resulting from the electron cyclotron maser instability occur at frequencies $f$ very near the fundamental electron cyclotron frequency (Treumann, 2006):

$$f = \frac{eB}{2\pi m_e}. \tag{8}$$

For magnetic field strengths $B$ in units of Gauss, the emission frequency in megahertz is approximated as $f_{[MHz]} \approx 2.8 B_{[Gauss]}$, and exoplanets with 10 Gauss fields such magnetic Jupiter analogs are expected to produce emissions at ∼28 MHz, while magnetic Earth analogs will be a factor of ∼ten weaker.

Above ∼10 MHz, astrophysical radio emissions including exoplanet aurorae can reach the ground. The Square Kilometer Array (SKA) is expected to attain sufficient sensitivity to detect some exoplanet radio aurorae from strongly magnetized gas giants (Grießmeier et al., 2007). However, the Earth's ionosphere is opaque to radio emissions at frequencies below ∼10 MHz. Detecting terrestrial exoplanet radio aurorae below this ionospheric cutoff will therefore require space-based arrays such as the Great Observatory at Long Wavelengths (GO-LoW; Knapp et al., 2024) and the lunar Farside Array for Radio Science Investigations of the Dark ages and Exoplanets (FARSIDE) (Burns et al., 2019).

The luminosities of auroral radio emissions depend in part on key magnetic properties of exoplanet systems. The power available to drive aurorae by the breakdown of plasma co-rotation scales with the planet's magnetic field $B_p$, its angular velocity $\Omega_p$, and its radius $R_p$ (Saur et al., 2021):

$$S \propto B_p^2\, \Omega_p^2\, R_p^2, \tag{9}$$

favoring objects with strong magnetic fields and rapid rotation, such as brown dwarfs. For numerical treatments of the magnetospheric-ionopsheric coupling currents that arise, we refer the interested reader to Nichols et al. (2012) and Turnpenney et al. (2017).

Power from aurorae driven by plasma flows scale with incident magnetic flux, which increases for close-in planets when all else is equal (Zarka et al., 2018). This motivates searches for radio aurorae from hot Jupiters, for which some dynamo theories predict stronger magnetic fields (e.g., Yadav & Thorngren, 2017). Stronger fields offer two advantages: they can lead to larger magnetospheric cross-sectional areas and therefore higher dissipated flux and expected auroral power, and they may produce aurorae at higher frequencies where today's low frequency arrays, such as LOFAR, are more sensitive.

However, no detections of exoplanet aurorae have been confirmed to date (Turner et al., 2021, and references therein), despite an initial promising detection of excess radio emission from the $\tau$ Boötis system at $21-30$ MHz with the beam-formed LOFAR array (Turner et al., 2021, 2023). Indeed, initial modeling finds that hot Jupiters may suffer from dense, plasma-filled magnetospheres that inhibit electron cyclotron maser emission, though instrument sensitivities remain a limiting factor.

Instead, aurorae have been conclusively detected on brown dwarfs and low mass M dwarfs, collectively known as ultracool dwarfs (Figure 5; e.g., Hallinan et al., 2007, 2008, 2015; Kao et al., 2016; Route & Wolszczan, 2012; Vedantham et al., 2020a, 2023, and references therein). Such objects can serve as laboratories for understanding extrasolar auroral physics (Pineda et al., 2017; Saur et al., 2021) as well as planetary dynamo models (Kao et al., 2016, 2018).

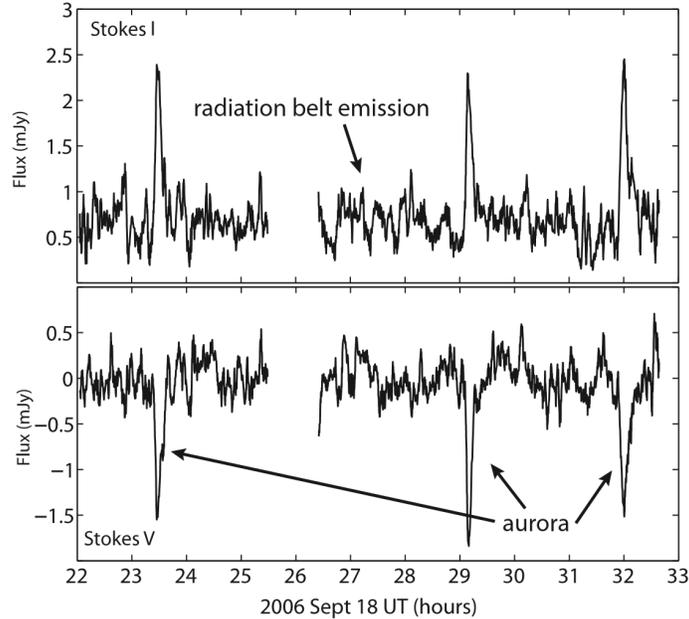

Figure 5: Stokes I and V radio timeseries at 8.44 GHz of the ultracool dwarf LSR J1835+3259 showing ~100% circularly polarized and rotationally periodic electron cyclotron maser bursts later identified as aurora by Hallinan et al. (2015). Using multi-epoch resolved imaging of this same object, Kao et al. (2023) showed that the non-auroral radio emissions are extrasolar Jovian radiation belt analogs (Radiation Belt Emissions subsection; see also Figure 8). Credit: adapted from Figure 1 of Hallinan et al. (2008) with permission.

While radio aurorae offer direct measurements of exoplanet magnetic fields, they probe only the local emitting region. As such, they provide lower bounds on a surface-averaged magnetic energy (Kao et al., 2016, 2018). Even so, these lower bound measurements can set stringent constraints on exoplanet dynamo models. Numerical dynamo simulations finding that convected energy flux sufficiently determines the magnetic field strengths of planets and brown dwarfs (Christensen et al., 2009) have been extended to models arguing that hot Jupiters may support strong magnetic fields (Yadav & Thorngren, 2017, see also discussion in the Star-Planet Interactions subsection). However, direct measurements of brown dwarf magnetic field strengths via their radio aurorae demonstrate otherwise[2] (Kao et al., 2016, 2018).

Obtaining topological information from aurorae requires dynamic spectra paired with careful modeling, as has been attempted for ultracool dwarfs with inconclusive results (Yu et al., 2011; Lynch et al., 2015). Furthermore, radio aurorae can be emitted along an object's magnetic field lines, resulting in broadband emission that increases with frequency as emitting regions approach strong magnetic fields close to the dynamo surface and then cut off sharply above an object's upper atmosphere at the high frequency end of the radio spectrum (Figure 6; Zarka, 2007). While the broadband nature of auroral radio emissions allow for relatively lenient frequency search spaces, obtaining the strongest constraints on surface-averaged magnetic field strengths requires broadband observations that can identify this cutoff frequency. However, care must be taken when interpreting non-detections, which can be attributed to many factors that are unrelated to objects' magnetic field strengths (Kao & Shkolnik, 2024).

---

[2] These models apply only to rapidly rotating objects with dipole-dominated magnetic fields (Christensen et al., 2009). Section 2 comments on the former requirement, while the confirmation of strong dipole fields traced by bright radiation belts around aurorae-emitting ultracool dwarfs (Kao et al., 2023; Climent et al., 2023) suggest that at least some aurorae emitting brown dwarfs may meet magnetic energy partition requirements.

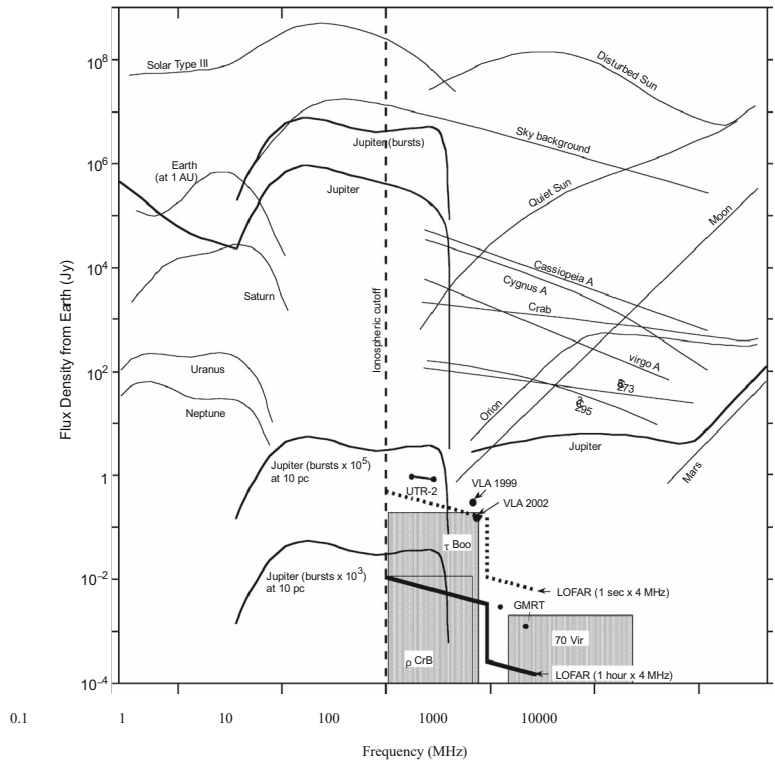

Figure 6: Radio spectra of auroral emissions from Solar System planets compared to other astrophysical radio sources and instruments. Aurorae are broadband and increase in frequency as magnetic field strengths increase near the emitting planet. Sharp cut-offs occur at the high frequency end of their spectra that correspond with emitting regions nearest the upper atmosphere of the planet. Credit: Zarka (2007).

A conclusive detection of radio aurorae from a free-floating exoplanet or planet-mass brown dwarf (e.g., Kao et al., 2016, 2018) requires demonstrating electron cyclotron maser emissions originating from the system that are periodic on the rotational timescale of the planet[3] or the orbital timescale of a suitable satellite (for the latter, see the Star-Planet Interactions subsection). Radio aurorae from an exoplanet that is gravitationally bound to its host star may exhibit similar behaviors, though the additional possibility of aurorae driven by stellar winds can relax phenomenological requirements: persistent electron cyclotron maser emissions that disappear during eclipse would also be sufficient. When combined with a broadband search for the auroral cutoff frequency, diagnosing exoplanet magnetic fields with radio aurorae can require significant observational investment, lending themselves well to large multi-frequency all-sky surveys.

*Helium 1083 nm transmission spectropolarimetry.* Recent theoretical developments of the detailed radiative transfer properties of helium absorption in escaping atmospheres (Oklopčić & Hirata, 2018; Oklopčić, 2019) have led to one of the most promising new methods for assessing gas giant magnetic fields: spectropolarimetric transit observations of He 1083 nm absorption in hot Jupiters (Oklopčić et al., 2020).

As helium atoms in the thermosphere and exosphere of a hot Jupiter absorbs background light from its host star, the intensity and spectrum of the incident stellar radiation can excite neutral helium to a $2^3S_1$ triplet state. This excited state is metastable because transitions to the ground state are exclusively highly forbidden, allowing sufficient populations of these excited helium atoms to accumulate and facilitating detections of absorption at He 1083 nm in a handful of transiting hot Jupiters (e.g., Spake et al., 2018; Allart et al., 2018, 2019; Nortmann et al., 2018; Salz et al., 2018; Alonso-Floriano et al., 2019).

---

[3] An object's rotation period measured from cloud variability can differ from the rotation period of its deep interior as measured from its radio aurorae. This effect has been observed on a cold T6.5 spectral type brown dwarf as well as on Jupiter (Allers et al., 2020, and references therein).

Polarization in He 1083 nm absorption features diagnoses exoplanet magnetic fields. The polarization arises from the Hanle effect, where atoms irradiated by incident linearly polarized light, such as from a stellar radiation field, preferentially align with and precess about an ambient magnetic field. Fine structure splitting of the He 1083 nm triplet then selectively absorbs specific linear polarizations of the incident light (Oklopčić et al., 2020). The resulting absorption feature exhibits circular or linear polarization (from the Zeeman and Hanle effects, respectively) depending on the orientation of the ambient magnetic field with respect to the observer's line of sight.

One advantage of applying the Hanle effect as a diagnostic of exoplanet magnetic fields is that it occurs for magnetic fields stronger than a few $\times\ 10^{-4}$ Gauss. In contrast, the Zeeman effect produces circular polarization but requires very strong magnetic fields $B > 3(\Delta v/10 \text{ km s}^{-1})$ kiloGauss to exceed the Doppler broadening of the line $\Delta v$ (~600 Gauss for HD 209458 b).

The linear polarization signal depends on optical depth and the line-of-sight components of the ambient magnetic field strength (Figure 7), with an expected signal between 0.1% and 0.01% depending on assumed geometries (Oklopčić et al., 2020). Accordingly, He 1083 nm spectropolarimetry favors systems with large transit depths and bright host stars.

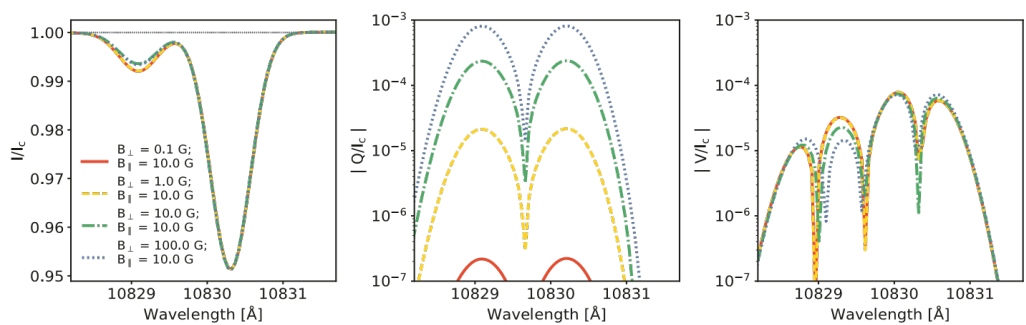

Figure 7: Line-of-sight magnetic field strengths impact the (left) total intensity profile of He 1083 nm absorption, (center) the linear polarization signal, and (right) circular polarization signal, where $I$, $Q$, $V$ are the corresponding Stokes parameters. Credit: Oklopčić et al. (2020).

*Radiation belt emission.* The previous two methods yield local (Helium 1083 nm transmission spectropolarimetry) or line-of-sight averaged magnetic field strengths (Exoplanet Aurorae). In contrast, the presence or lack of radiation belt emission can directly assess an object's magnetic topology.

All of the Solar System planets with large-scale magnetic fields maintain high energy plasma populations, with energies of tens of MeV, in the equatorial regions of their magnetic dipole components (Figure 8; Mauk & Fox, 2010). Electrons populating planetary radiation belts produce radio emissions that differ from aurorae: instead of coherent electron cyclotron maser emissions near the fundamental cyclotron frequency, Solar System radiation belts produce incoherent synchrotron emissions (e.g., Bolton et al., 2002), which occur at hundreds to thousands of times the local electron cyclotron frequency depending on emitting electron energies (Rybicki & Lightman, 1986). Radiation belt emissions vary more gradually than auroral emissions (Kao et al., 2023), so they are particularly well-suited to snapshot searches (e.g Kao et al., 2019; Richey-Yowell et al., 2020; Kao & Pineda, 2022). Finally, their broadband emission at high cyclotron frequency harmonics allows the use of radio arrays tuned to a broad range of frequencies that can exceed the fundamental cyclotron frequencies of exoplanet magnetic fields.

In addition to Earth, Jupiter, Saturn, Uranus, and Neptune (Mauk & Fox, 2010), all brown dwarfs that produce radio aurorae at gigahertz frequencies also exhibit non-auroral radio emissions (Pineda et al., 2017; Kao et al., 2019; Kao & Shkolnik, 2024, and references therein). Such emissions can trace extrasolar analogs to Jovian radiation belts (Figure 8; Kao et al., 2023; Climent et al., 2023).

Solar system and brown dwarf radiation belts extend well beyond several times the object's radius (Bolton et al., 2004; Kollmann et al., 2018; Kao et al., 2023; Climent et al., 2023). Despite these extended sizes, resolved radiation belt imaging will be limited to all but the nearest exoplanets until space-based radio arrays can offer baselines exceeding the diameter of Earth. However, confirming that the quasi-quiescent component of brown dwarf radio emissions traces their radiation belts opens a pathway to identifying extrasolar radiation belts using *unresolved* radio timeseries with future arrays. For exoplanets bound to their host stars, distinguishing between stellar and exoplanet radio emissions in unresolved radio timeseries will add an additional layer of difficulty. Free-floating planets are untroubled by this concern, as two isolated planetary mass brown dwarfs exhibiting radiation belt emissions showcase (Kao et al., 2016).

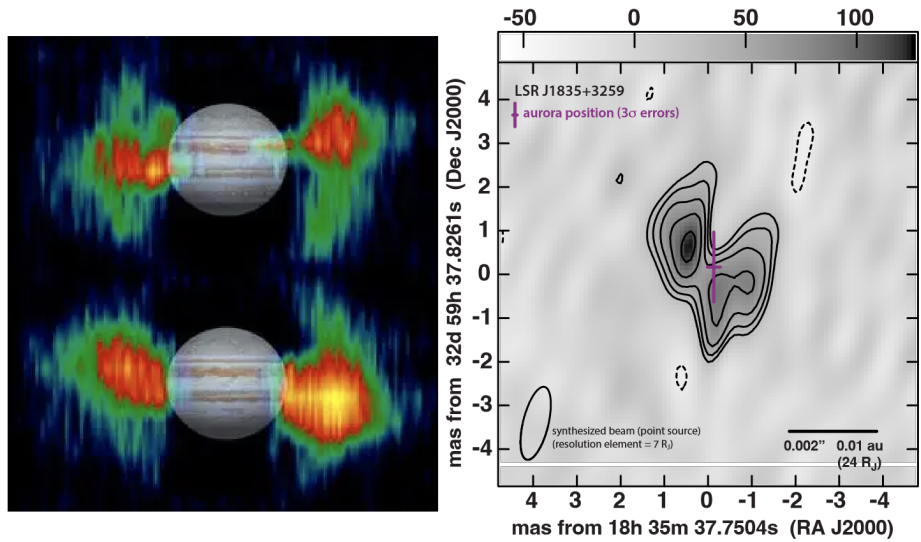

Figure 8: (left) 2 GHz imaging of Jupiter's inner electron radiation belts at different rotational phases. Jupiter's dipole moment is offset from its rotation axis. Credit: NASA/JPL. (right) 8.4 GHz imaging of an electron radiation belt around the aurora-emitting ultracool dwarf LSR J1835+3259. Credit: Kao et al. (2023).

At present, topological information from detections of extrasolar radiation belts is limited: they confirm only that an object has a dipole field component but offer no further information on the partition of magnetic energies between dipole and higher-order field components. For instance, Uranus and Neptune host radiation belts (Mauk & Fox, 2010) though dipoles do not dominate their magnetic fields (Stevenson, 2010). In contrast, Mars and Venus do not have global magnetic fields or radiation belts.

Finally, brown dwarf radiation belt emissions are fainter than radio aurorae by factors of at least a few (Kao et al., 2016, 2018, 2023; Pineda et al., 2017, and references therein). If exoplanet radiation belt emissions scale similarly, they will likely be too faint to be detectable with current or even next generation instruments like the ngVLA, SKA, and DSA 2000. If astronomers wish to access the model-independent topological information offered by radiation belts, investing in future generations of high sensitivity radio arrays with very long baseline interferometry capabilities will be necessary.

**Indirect Inferences**

*Star-Planet Interactions.* Close-in companions can excite aurorae on their hosts, which offer model dependent means of indirectly inferring basic magnetospheric properties of the companions. In the Solar System, companion-driven aurorae occur on Jupiter (Figure 9; see also Mura et al., 2017) and scaling up theory grounded in Solar System objects suggests that they can also occur on host stars with close-in planets. In particular, magnetized and/or ionized companions must reside within their hosts' sub-Alfvénic magnetospheric regions, such that hostward-traveling Alfvén waves perturbed by these companions can exceed companion-ward plasma velocities to reach the host (Saur et al., 2013, and references therein).

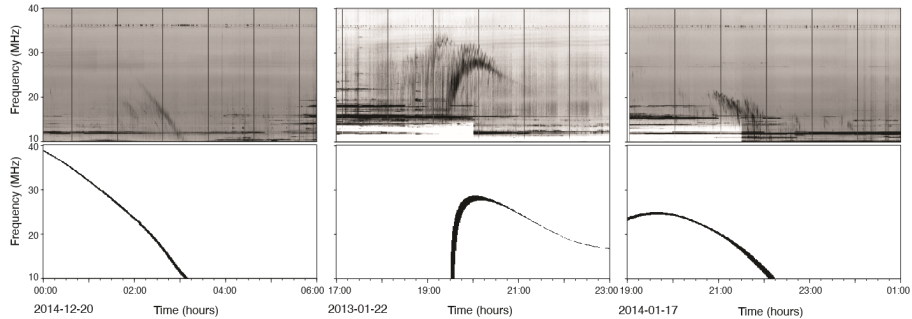

Figure 9: Io-induced radio auorae from Jupiter and modeling (top, bottom). Credit: adapted from Figure 5 in Louis et al. (2019) under the terms of the Creative Commons Attribution License.

In sub-Alfvénic interactions, Poynting flux (electromagnetic energy flux) can radiate away from the companion toward the host. The total dissipated power $S$ in companion-driven aurorae depends on the host star's wind properties and the obstacle size presented by the companion, $R_o$:

$$S \propto R_0^2 \, B_{\text{wind}} \, \Delta u^2 \, \sin^2 \theta \, \sqrt{\rho_{\text{wind}}}, \tag{10}$$

where $B_{\text{wind}}$ is the host star's magnetic field at the planet's location, $\Delta u$ is the apparent wind velocity with respect to the companion, $\theta$ is the angle between the stellar magnetic field and the apparent wind velocity, and $\rho_{\text{wind}}$ is the wind mass density (Figure 10; Saur et al., 2013). The obstacle size folds in the companion's magnetic field. All other factors being equal, a companion with a large magnetosphere will present as a larger obstacle than one possessing only an ionosphere, and it will consequently dissipate more power. This model, known as the Alfvén wing model, can produce flux values that are consistent with aurorae driven by Io, Europa, Ganymede and Callisto on Jupiter (Saur et al., 2013). However, it has not yet been validated for any exoplanet system.

Crucially, star-planet interaction radio emissions induced on the host will occur at frequencies corresponding to the host's magnetic field. This offers an incredible advantage: although radio aurorae from terrestrial planets cannot be observed from the ground (See the Exoplanet aurorae subsection above), magnetized terrestrial planets may induce aurorae on their more strongly magnetized hosts, which can produce radio emissions at frequencies that are accessible today.

However, diagnosing exoplanet magnetic properties with star-planet interaction emissions requires addressing several challenges. First, one must demonstrate that the observed emission is unequivocally attributable to a star-planet interaction by showing that it is periodic on the orbital timescale of the responsible planet and cannot be associated with other stellar magnetic activity. Second, one must understand how the total dissipated power partitions into emissions at different wavelengths to convert $S$ to the auroral power observed at the wavelength(s) of interest. Third, one must have a reasonably accurate stellar wind model. Fourth, one must know the ephemerides of the companion. Even with all this information in hand, one can only infer an obstacle size. Obtaining a magnetospheric size requires comparing this obstacle size to a well-characterized planet radius (see the Ohmic Dissipation subsection below for additional considerations), such as one derived from transit data. Finally, one will need to account for the impact of the host star's plasma pressure on the magnetopause of the planet.

As recent radio detections of candidate star-planet interaction emissions demonstrate, the first requirement necessitates significant investment of observing resources as well as a mature understanding of stellar flare behaviors. Distinguishing auroral radio emissions from plasma emissions can be difficult, and we refer the reader to Villadsen & Hallinan (2019) and Vedantham et al. (2020b) for in-depth discussions. Additionally, follow-up radio campaigns have not yielded a repeating detection of 150 MHz emission from the M dwarf GJ 1151 that initial modeling suggested a hypothetical close-in terrestrial companion could plausibly excite (Vedantham et al., 2020b). Radial velocity campaigns rule out massive close-in companions around this star (Pope et al., 2020; Blanco-Pozo et al., 2023) but cannot rule-in a terrestrial companion. Pineda & Villadsen (2023) reported repeating candidate detections of 2-4 GHz star-planet interaction emission from the known planet host YZ Ceti, but the emissions exhibit some scatter when phase-folded with the orbital period of YZ Ceti b. System geometries may explain this scatter (Pineda et al., 2017; Trigilio et al., 2023), but such effects are not yet well understood (Kavanagh & Vedantham, 2023). Finally, ordinary stellar magnetic activity processes can produce radio emissions that may masquerade as star-planet interactions (Villadsen & Hallinan, 2019; Yu et al., 2023) or other radio emissions interpreted as aurorae on active stars (Zic et al., 2019; Villadsen & Hallinan, 2019). Here, the old adage "know thy star, know thy planet" strikes true.

Brown dwarfs again may offer a laboratory for testing star-planet interaction models. As close-in companions around brown dwarfs are discovered, we may find that some drive radio aurorae on their hosts. Companion interactions show uniquely characteristic arcs in the dynamic spectra (time- and frequencydependent flux densities) of radio aurorae (Figure 9). Detailed models of these radio arcs can yield information about the orbits, rotation, and magnetic fields of companions (Hess et al., 2008). Similar models are being developed for future detections of exoplanet radio emission.

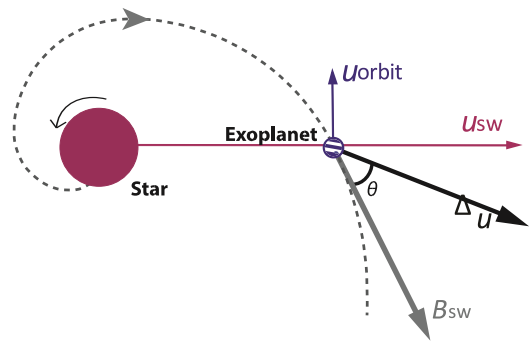

Figure 10: Schematic of quantities relevant to the Alfvén wing model of star-planet interaction by Saur et al. (2013) and summarized in Equation 10. The dashed line represents a Parker spiral stellar magnetic field. Credit: adapted from Figure 1 of Turnpenney et al. (2018).

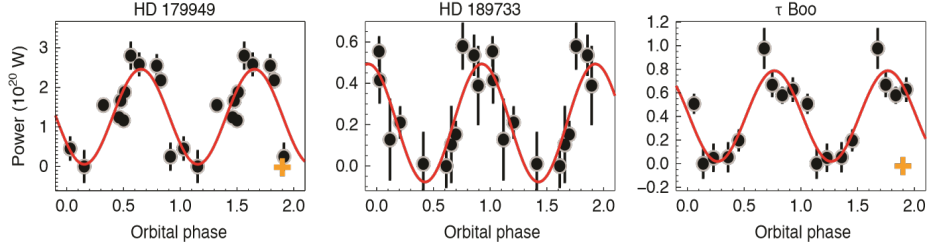

Figure 11: Orbitally modulated Ca II K emission attributed to star-planet interactions from hot Jupiter systems. Credit: adapted from Figure 1 in Cauley et al. (2019).

This section would be incomplete without a discussion of orbitally modulated magnetically active stellar chromospheric lines. The prevailing interpretation for such emissions are star-planet magnetospheric interactions and they have been observed from hot Jupiter systems including HD 179949 b, HD 189733 b, τ Boötis, and υ Andromedae (Figure 11 Shkolnik et al., 2003, 2005, 2008; Gurdemir et al., 2012; Cauley et al., 2018, and references therein). Several theories have been proposed to explain the power observed in these emissions, including reconnection events between stellar and planetary magnetic field lines as the latter travels through the stellar magnetosphere (Cuntz et al., 2000), reconnection events triggered by the planet on the stellar surface (Lanza, 2009, 2012), the Alfvén wing model (Saur et al., 2013), and the stretching of magnetic flux tubes between the star and the planet as the latter orbits (Lanza, 2013).

Of these, only the last model can account for the total power *P* derived from detections of star-planet chromospheric emissions from hot Jupiters:

$$P \approx \frac{2\pi}{\mu} f_p R_p^2 B_p^2 \text{v}_{rel}, \tag{11}$$

where $f_p$ is the fractional hemispheric flux tube coverage for the planet and depends on the ratio between the planet's surface polar magnetic field $B_p$ and the stellar magnetic field at the location of the planet, $R_p$ is the planet radius, and $v_{rel}$ is the relative velocity between the planet and stellar magnetic field lines at the planet's location. Although this model can explain observations, the magnetic field strengths that Cauley et al. (2019) derive using it remain unconfirmed by an independent means. As with the Alfvén wing model, this model also relies on an accurate understanding of energy partitions; inferred exoplanet magnetic fields can differ by several orders of magnitude depending on the fraction of the total star-planet interaction power that is radiated in the observed chromospheric lines. Tuning this energy partition allow Cauley et al. (2019) to infer magnetic field strengths for HD 179949 b, HD 189733 b, τ Boötis, and υ Andromedae that agree with predictions from magnetic dynamo scaling relationships relying primarily on a planet's convected thermal energy (Christensen et al., 2009; Yadav & Thorngren, 2017). However, direct measurements of brown dwarf magnetic field strengths call this tuning into question (Kao et al., 2018), underscoring the necessity of independently validating these inferred magnetic field strengths with other measurement methods and/or detailed energy partition studies.

***Ohmic Dissipation.*** Above 1000 K equilibrium temperatures, many hot Jupiters exhibit inflated radii relative to predictions from planetary evolution models (e.g., Thorngren & Fortney, 2018). One proposed explanation for this "radius anomaly" is heating from interior Ohmic dissipation in electrical currents induced by ionized winds in magnetized planets (Batygin & Stevenson, 2010; Batygin et al., 2011).

Ohmic dissipation cannot account for all observed radius anomaly behaviors (Wu & Lithwick, 2013; Ginzburg & Sari, 2016), and interior heating may also arise from tidal dissipation and shear instabilities or vertical mixing. As such, inferring inflated hot Jupiter magnetic fields from Ohmic dissipation models (e.g., Wu & Lithwick, 2013; Rauscher & Menou, 2013) requires carefully accounting for all such interior heating processes. We refer the interested reader to Fortney et al. (2021) for additional discussion on hot Jupiter interior heating mechanisms.

*Magnetospheric bow shocks.* Stellar winds interacting with an exoplanet's magnetic field can form bow shocks when they reach supersonic speeds and are observed around Solar System planets. Under certain conditions, these magnetospheric shocks may absorb excess stellar radiation ahead of a planet's orbit, resulting in both deeper transits and earlier transit ingresses, and perhaps even pre-transit absorption (Figure 12), at wavelengths where the shock is optically thick (e.g., Vidotto et al., 2010).

Magnetospheric stand-off distances inferred from pre-transit absorption can then place lower-bound limits on exoplanet magnetic field strengths by balancing the stellar wind magnetic pressure $B_w$ at the location of the planet against the pressure of the planet's magnetic field $B_p$ at its magnetopause and neglecting ram and thermal pressure terms (Vidotto et al., 2010):

$$\frac{B_w^2}{8\pi} + \rho_w \Delta u_w^2 + p_w = \frac{B_p^2}{8\pi} + p_p. \tag{12}$$

Magnetospheric bow shocks are only one of several proposed interpretations (Lai et al., 2010; Vidotto et al., 2010; Llama et al., 2011, 2013) for enhanced UV and/or Balmer line transit depths and early ingress detections compared to optical transits observed on WASP-12b (Fossati et al., 2010; Haswell et al., 2012) and HD 189733b (e.g., Ben-Jaffel & Ballester, 2013; Bourrier et al., 2013; Cauley et al., 2015, 2016). More detailed numerical hydrodynamical simulations find optical depths requiring rapid cooling of the shocked wind that is difficult to achieve (Alexander et al., 2016), casting doubt on but not ruling out the magnetospheric bow shock interpretation for WASP-12b. Other interpretations, such as overflow of the planet's Roche lobe (Lai et al., 2010; Bisikalo et al., 2013) and charge exchange between stellar winds and planetary outflows (Tremblin & Chiang, 2013), can also reasonably but incompletely reproduce observed transit features (Cauley et al., 2016, and references therein).

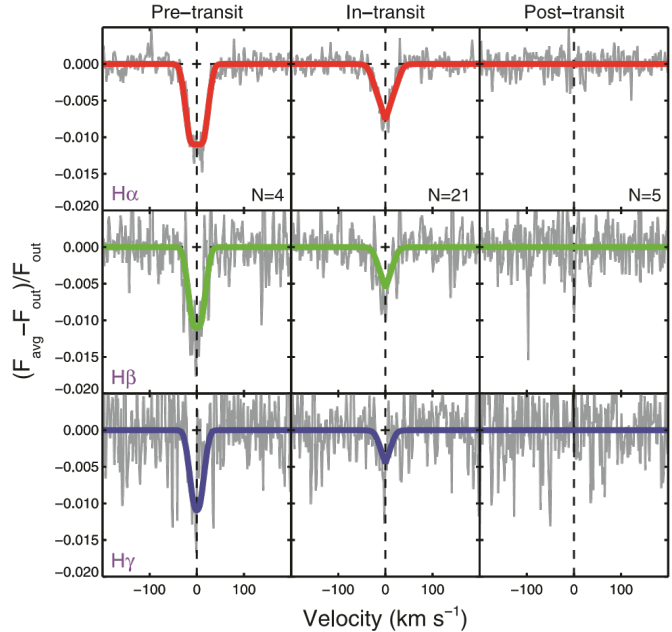

Figure 12: Pre-transit Balmer line absorption from averaged transmission spectra of the hot Jupiter HD 189733b (grey: averaged data; red/green/blue: averaged model), initially attributed as a thin bow-shock though later modeling cast this interpretation into doubt. Credit: adapted from Figure 1 in Cauley et al. (2015).

*Atmospheric outflow transit spectroscopy.* Stellar radiation can drive highly ionized outflowing winds on close-in planets (See the Atmospheric Escape subsection below). In regions of a planet's magnetosphere where magnetic pressure dominates thermal and ram pressures from its photoevaporative winds, atmospheric outflow is effectively "anchored" along the planet's magnetic field lines. In contrast to the magnetically confined strong field regime, in the unconfined weak field regime where thermal and ram pressures dominate, day-to-night winds can occur instead (e.g., Batygin et al., 2013; Adams & Owen, 2015).

Three dimensional magnetohydrodynamic modeling that calculates magnetic drag from local conditions – most importantly, temperature – find that highly irradiated ultra hot Jupiters exhibit magnetically sensitive circulation patterns in their upper atmospheres. Neglecting a localized treatment of magnetic drag results in day-to-night winds that observationally manifest as blue shifted transmission spectra. In contrast, the introduction of even a weak magnetic field can shift dayside winds poleward and introduce sufficient magnetic drag to somewhat suppress blueshifts and/or introduce mid-transit redshifts for certain species (Beltz et al., 2023). Magnetic drag may also influence emission spectra and phase curves (Beltz et al., 2022a,b) and in some cases magnetic models fit phase curves better than the non-magnetic model (Coulombe et al., 2023). Similarly, two-dimensional modeling finds blue-shifted He 1083 nm transits in the unconfined wind regime that do not occur when planet winds are magnetically confined (Figure 13). Instead, these transits are deeper than in the magnetically unconfined regime, though many parameters can influence transit depths (Schreyer et al., 2024).

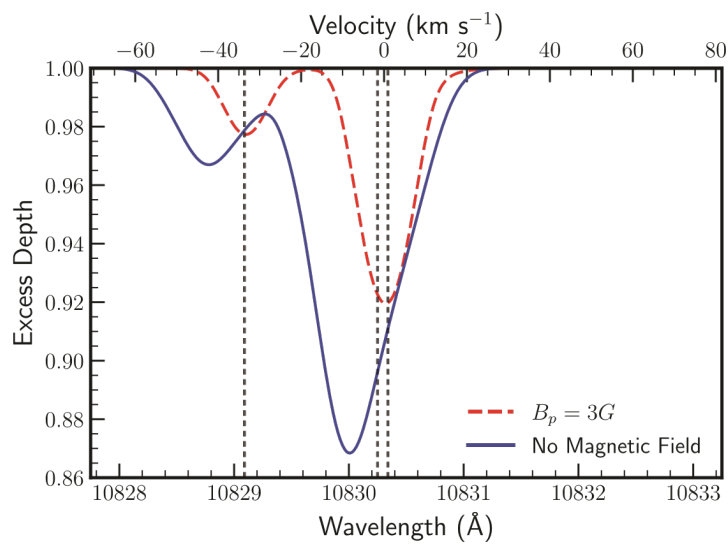

Figure 13: Modeled He 1083 nm absorption for a magnetized and unmagnetized transiting hot Jupiter. In the unmagnetized case, day-to-night winds result in a net blueshift. Credit: Figure 4 from Schreyer et al. (2024).

Comparing He 1083 nm transit spectroscopy to models can, in theory, distinguish between the confined and unconfined regimes, but diagnosing actual exoplanet magnetic field strengths requires more careful interpretation. In particular, the confined and unconfined regimes depend on the magnetic strength, topology, and orientation of the planet, as well as the stellar wind environment and ionizing flux. Additionally, the orbital eccentricity must be well known, as eccentric orbits introduce line-of-sight velocity shifts that linearly depend on the eccentricity deviation from zero. For typical eccentricity uncertainties of order ∼0.1, the resulting velocity shift is comparable to the effect of magnetically sensitive outflow geometries (Schreyer et al., 2024). Finally, as is the case for other indirect methods discussed in this chapter, independent cross-validation of these modeled effects with other measurements of exoplanet magnetic fields remains to be seen.

**A promising future**

The reader may notice that no exoplanet magnetic field strengths are quoted in this section. This choice is deliberate, as no confirmed direct measurements of exoplanet magnetic field strengths exist at the time of writing, reflecting the aspirational nature of most of the discussed methods and the instrumentation and interpretative challenges that they must overcome. Nevertheless, tantalizing observational hints of exoplanet magnetism have inspired exciting theoretical developments and showcase a future ripe with possibility. Even as we look toward the forthcoming era of exoplanet magnetic field measurements, we urge readers to look to brown dwarfs today as laboratories that can test and refine models of aurorae, radiation belts, and star-planet interactions – and the information we may learn from them in the future when they are detected in exoplanet systems.

# INFLUENCES OF EXOPLANET MAGNETIC FIELDS

Knowing whether an exoplanet possesses a magnetic field can influence our knowledge about other important properties of the planet and the processes that occur there. Perhaps the most obvious information comes about the interior of the planet (as discussed in the Theory of Exoplanet Magnetic Dynamos section, including its internal heat and its potential, in the case of rocky exoplanets, for geologic activity). But other important information can be inferred about: (1) the manner in which the planet interacts with its space environment and the extent of this interaction region; (2) whether the planet is likely to retain an atmosphere; (3) particle precipitation into the planet's atmosphere and/or surface and its consequences; and (4) the evolution of the planet's rotation rate and orbit. We discuss each of these four areas in the sections below.

**Interaction with Space Environment**

The nature of a planet's magnetic field determines how it interacts with its space environment. Planetary bodies possessing internally generated magnetic fields form *intrinsic* magnetospheres, where incident flowing plasma (i.e., the stellar wind) is deflected around the object by the planet's field. In our own solar system, Mercury, Earth, Jupiter, Saturn, Neptune, and Uranus all have intrinsic magnetospheres, along with Jupiter's moon Ganymede (see Table 1). Planetary bodies possessing a sufficiently conducting layer (e.g., an ionosphere or a salty ocean) may form *induced* magnetospheres, where the incident flowing plasma induces a current in the conductors, which in turn generates a magnetic field that deflects the incident plasma. Venus, Mars, Pluto, comets, Jupiter's moon Europa, and Saturn's moon Titan all have induced magnetospheres. Other objects, such as Earth's Moon, have no significant internally generated magnetic field and no significant conducting layer, and incident plasma strikes the surface directly.

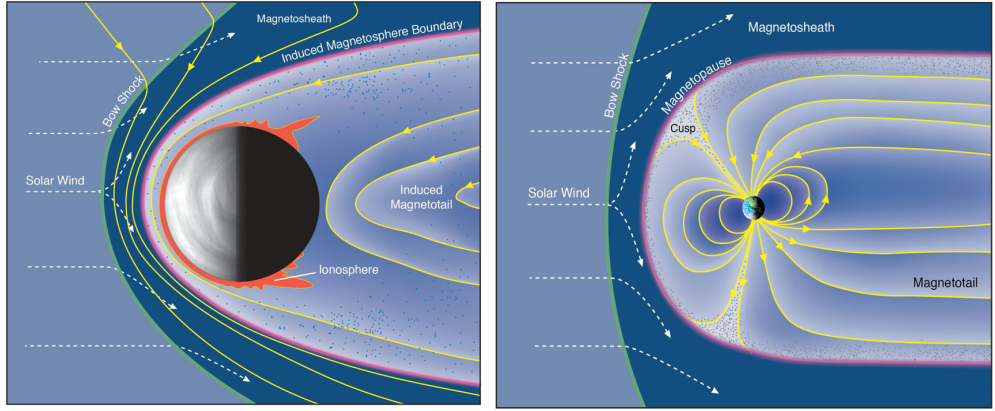

Figure 14: Schematics of characteristic induced (top) and intrinsic (bottom) planetary magnetospheres. Induced magnetospheres tend to be smaller relative to the size of the planet. Courtesy S. Bartlett, from Brain et al. (2013).

Induced magnetospheres are typically smaller than intrinsic magnetospheres relative to the size of the planet (Figure 14). One metric for magnetospheric size is the 'standoff distance' of the solar wind, determined to first order through considering the balance between the pressure from the incident stellar wind ($\rho_{SW} v_{SW}^2$) and the magnetic pressure generated at the planet ($B_P^2/(2\mu_0)$). The standoff distances for Earth, Jupiter, and Neptune are about 10, 50, and 25 planetary radii respectively (see Kivelson & Bagenal, 2007), while Venus, Mars, and Pluto have standoff distances of about 1.05, 1.3, and 3 planetary radii (see Brain et al., 2016). This pattern does not always hold; in some cases, an intrinsic planetary field can be relatively weak (e.g., Mercury, with a small intrinsic magnetosphere) or an induced planetary field can be relatively strong or be complemented by dynamic ($\rho_P v_P^2$) and thermal plasma (*nkT*) pressure from the planetary object (e.g., active comets, with large induced magnetospheres). But given the same external conditions and no significant planetary atmospheric outflow we generally expect intrinsic magnetospheres to be larger than induced magnetospheres. Therefore, exoplanets without outflows and with internally generated fields should influence a larger region of space than if they were unmagnetized.

The presence and nature of a significant internally generated magnetic field also partially determines the processes that occur near a planet and in its upper atmosphere. Intrinsic magnetospheres may trap radiation belts similar to those at Earth and Jupiter, allowing the magnetic field to be detected (see the Radiation Belt emission subsection above). Aurorae are not unique to intrinsic magnetospheres (Schneider et al., 2015). However, aurorae in intrinsic magnetospheres are usually spatially confined to regions near the magnetic poles and are brighter (in our own solar system) than the more global aurora that occur in induced magnetospheres. Uranus and Neptune possess magnetic fields that have significant non-dipolar components and are highly titled relative to the planet's rotation axis (Ness et al., 1986; Connerney et al., 1991). Both features have consequences for the access of stellar wind particles into the magnetosphere. Close-in magnetized exoplanets may be directly connected to the stellar magnetic field, resulting in a so-called star-planet interaction (SPI) that enables direct exchange of plasma along open magnetic flux tubes from the planet (see the Star-Planet Interactions subsection above).

**Atmospheric Escape**

A possible consequence of a planet possessing a magnetic field is the influence of the magnetic field on the escape of atmospheric particles. This topic is much debated in recent years. On one hand, it has often been thought that planetary magnetic fields act as shields for their atmospheres, preventing stellar winds from accessing the atmosphere and driving escape (e.g., Hutchins et al., 1997; Lundin et al., 2007; Dehant et al., 2007). On the other hand, atmospheric escape does not necessarily require direct access of a stellar wind to a planet's atmosphere; energy can transfer to atmospheric atoms and molecules via a variety of processes, and a stellar wind can transfer energy to an atmosphere indirectly (Moore & Horwitz, 2007; Brain et al., 2013; Tarduno et al., 2014; Brain et al., 2016; Del Genio et al., 2020; Gronoff et al., 2020; Maggiolo & Gunell, 2021; Brain, 2021; Vidotto, 2021). Whether (and how) a planet's magnetic field influences its escape rate likely depends upon many characteristics of the planet and its host star. Additionally, the influence on escape rates depends upon which escape process and which atmospheric species is being considered.

*Hydrodynamic Escape.* At present most analyses of atmospheric escape from exoplanets consider hydrodynamic escape, sometimes called photoevaporation. Hydrodynamic escape occurs in the fluid limit of Jeans (thermal) escape, when a fraction of the thermal distribution of atmospheric particles has energies that exceed the energy required to escape from the planet. When this fraction becomes significant, the escaping flow can be approximated as a fluid. A recent review of hydrodynamic escape from exoplanets was provided by Owen (2019). An estimate of the hydrodynamic loss rate for planets, Φ, can be obtained if one has estimates for the planet's mass ($M_p$) and size ($R_p$) as well as the flux at EUV wavelengths (∼ 10-120 nm) from the star ($F_{EUV}$), the radius where the EUV fluxes are absorbed ($R_{EUV}^2$), and some efficiency factor ($\epsilon$) for the conversion of incident EUV energy into atmospheric particle heating (e.g., Erkaev et al., 2007):

$$\Phi = \frac{\varepsilon\, \pi\, F_{EUV}\, R_{EUV}^2\, R_P}{G\, M_P}. \qquad (13)$$

This equation represents the "energy-limited" hydrodynamic escape rate, where the available incident EUV energy (as opposed to supply of atmospheric particles from the lower atmosphere) controls the escaping particle flux.

Hydrodynamic escape most likely occurs when the planet's atmosphere contains light species (such as hydrogen), the gravity of the planet is weak, or the stellar EUV flux is large (as is the case for active or young stars). For example, hydrodynamic escape is thought to have stripped the primordial H/He atmospheres of the terrestrial planets early in solar system history, when stellar EUV fluxes were considerably larger (Watson et al., 1981). But the reduced EUV flux experienced by the giant planets and their deep gravitational potentials have allowed them to retain significant H/He atmospheres. Hydrodynamic escape may also be responsible for a dearth of close-in exoplanets intermediate in size between Earth and Neptune (e.g., Fulton et al., 2017). Given the above, the focus on hydrodynamic escape over other escape mechanisms reflects current transit spectroscopy, often of exoplanets orbiting active stars at close distances). To first order, hydrodynamic escape is a thermal process driven by stellar EUV fluxes, resulting in the loss of predominantly neutral atmospheric particles. Planetary magnetic fields might thus be considered to have little influence on hydrodynamic escape rates. However, recent studies have considered the idea that outflowing hydrodynamic winds from planets undergo ionization from the same intense stellar EUV fluxes that heated them initially (e.g., Trammell et al., 2011; Kislyakova et al., 2014; Khodachenko et al., 2015; Erkaev et al., 2017; Owen & Adams, 2019). Once ionized, the outflowing particles can interact with a planet's magnetic field. Atmospheric particles that are ionized on closed planetary magnetic field lines may be trapped and re-impact the atmosphere, while those on open field lines are likely to escape the planet but may be reconfigured into a magnetotail shape. The orientation of the magnetic field with respect to the field carried by the stellar wind is therefore important since it influences the topology (open vs. closed nature) of fields close to the planet.

Overall, however, one expects hydrodynamic escape rates to be reduced if a planetary magnetic field is present, and the fraction of open magnetic field lines (approximated by the extent of the polar cap) provides an important metric for this attenuation (e.g., Trammell et al., 2011; Khodachenko et al., 2015). The conversion of escaping neutral hydrogen to protons should modify the shape of Lyman-alpha absorption lines in the spectrum of a transiting planet. Several studies (e.g., Kislyakova et al., 2014; Erkaev et al., 2017; Villarreal D'Angelo et al., 2018; Ben-Jaffel et al., 2022) have suggested that the shape of the line can couple with models for the stellar wind interaction with a magnetized planet to infer both the outflow rate and the strength of the magnetic field (e.g., Figure 15), as discussed in the Atmospheric Outflow Transit Spectroscopy subsection above.

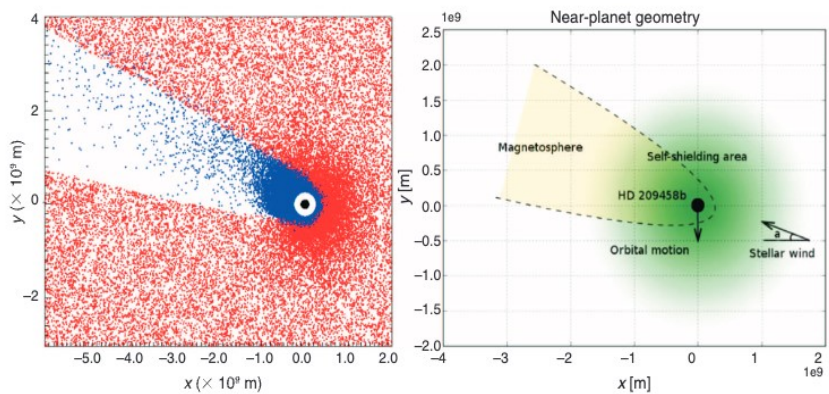

Figure 15: (Left) The simulated hydrogen exosphere around transiting exoplanet HD 209458b, with neutral atoms indicated in blue and ions in red. (Right) Cartoon showing the size and orientation of the planet's magnetosphere in the simulation. The situation is consistent with observed absorption in the red and blue wings of stellar Lyman-alpha during transit periods. From Kislyakova et al. (2014).

*Non-thermal Escape.* Planets that are not experiencing hydrodynamic escape through some combination of larger size, sufficiently small stellar EUV flux, and more massive atmospheric species characteristic of a secondary atmosphere are still subject to classic Jeans escape and to non-thermal escape processes including ion escape, photochemical escape, and sputtering. Ion escape results when charged particles are accelerated (via a combination of electric fields) to the energy required for escape. Photochemical escape occurs when exothermic chemical reactions are driven in the planet's upper atmosphere, usually by solar UV fluxes and often involving the dissociative recombination of molecular ions. Sputtering occurs when particles incident upon an atmosphere (either stellar wind particles or escaping atmospheric ions) collide with target atmospheric particles, "splashing" them out of the atmosphere. A notable aspect of non-thermal escape processes is that they can be more effective at removing massive atmospheric species (e.g., oxygen, carbon, nitrogen, noble gases) than thermal escape. Thus, they are important to evaluate, especially for planets possessing secondary atmospheres.

A planetary magnetic field can influence all three of the non-thermal processes identified above, since all three involve charged particles - and these influences may be important (e.g., Airapetian et al., 2017). The influence of a planetary magnetic field on photochemical escape is likely to be small. This partly because the energy that results from photochemical reactions may not be sufficient for escape on large planets where global magnetic fields are more likely. Sputtering may be more strongly influenced by the presence of a magnetic field, which deflects the stellar wind at larger distances from the planet and reduces the flux of stellar wind particles that could cause sputtering.

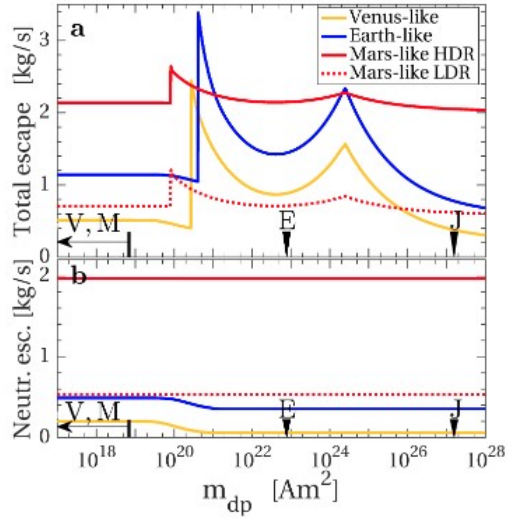

Figure 16: Semi-analytic model of atmospheric escape rates from Venus-, Earth-, and Mars-like planets as a function of planetary magnetic field strength. Calculations for Mars are performed for high and low dissociative recombination rates of molecular oxygen ions. the current dipole field strengths of Venus, Mars, Earth, and Jupiter are indicated. From Gunell et al. (2018).

The connection of planetary magnetic fields to ion escape is the most direct and is therefore the most often considered. Multiple studies have considered the influence of a magnetic field on ion escape (and sometimes other non-thermal processes). Several studies have taken semi-analytic approaches. Most do not consider exoplanets specifically but are either general or generalize from solar system planets. For example, Kulikov et al. (2007) examined the influence of an early Martian magnetic field on escape, concluding that the presence of a magnetic field likely decreased escape rates due to the larger standoff distance of the magnetic field. Blackman & Tarduno (2018) used a purely theoretical approach to consider the increased capture of energy from the stellar wind by a planet's magnetic field (compared to the case with no field), the slowing of the incident solar wind by the planetary field, and the recapture of escaping particles. They concluded that the impact of a planetary magnetic field on atmospheric escape depends upon the details of each of these three processes. Gunell et al. (2018) constructed semi-analytic models for several escape processes and applied them to Venus, Earth, and Mars (Figure 16). They found that the total atmospheric escape rate is comparable for the three planets and inferred that the presence of a planetary magnetic field does not reduce atmospheric escape. Finally, Ramstad & Barabash (2021) provided a formalism for considering atmospheric escape from many planets that accounts for various processes that can inhibit the escape of ions: the supply of ions from the lower atmosphere, the energy transfer from the solar wind to ions, and the efficiency of transport of ions from the upper atmosphere to space. They concluded that there is no solid evidence that planetary magnetic fields reduce atmospheric ion escape. Instead, they may increase escape rates.

There is observational support for the idea that planetary magnetic fields do not necessarily reduce atmospheric escape rates. In situ spacecraft observations of ion escape from terrestrial planets show that the escape rates from Venus, Earth, and Mars agree to within a factor of a few (Strangeway et al., 2010; Barabash, 2010; Ramstad & Barabash, 2021), with *greater* ion escape fluxes from Earth. There is some question about what fraction of the measured ion fluxes at Earth actually escape the planet (similar to the discussion of hydrodynamic escape above), since Earth's magnetic field may redirect upward-moving atmospheric ions measured in the magnetospheric polar regions back to the planet (Seki et al., 2001). However, recent measurements of ion fluxes in different regions of Earth's magnetosphere suggest that a significant fraction (>50%) escape (Slapak et al., 2017). Earth also appears to be more responsive to changes in the solar wind pressure than either Venus or Mars (Ramstad & Barabash, 2021).

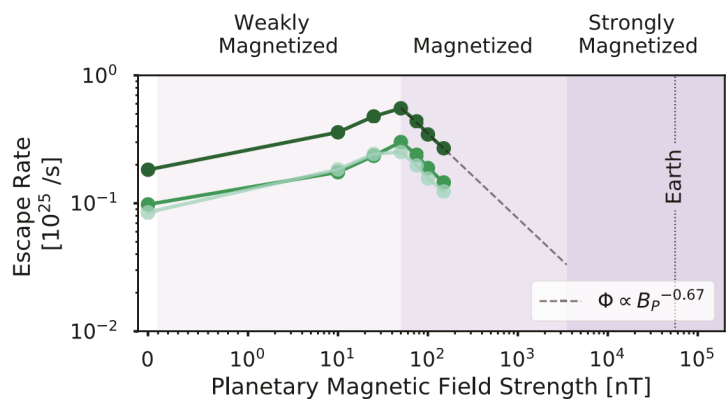

Figure 17: Ion escape rates increase with magnetic field strength before quenching for strongly magnetized planets. The dashed line indicates an extrapolated escape rate ($\Phi$) based on the simulations. The vertical dotted line indicates Earth's current dipole field strength. From Egan et al. (2019).

Detailed plasma simulations of the interaction of a stellar wind with a planetary atmosphere (e.g., Figure 17) suggest that there is not a monotonic relationship between the strength of a planet's magnetic field and its escape rate (Egan et al., 2019; Sakata et al., 2020, 2022). Instead, there is an intermediate planetary magnetic field strength for which ion escape rates are maximized. At lower planetary field strengths, a significant fraction of newly born ions is carried back into the planet's atmosphere by the incoming stellar wind. At higher field strengths, newly born ions are increasingly likely to be trapped in a large planetary magnetic field, unable to escape.

A few studies have considered non-thermal ion escape and related processes from specific exoplanetary systems. Cohen et al. (2014) explored Earth-like planets orbiting M Dwarf stars, and noted that potentially habitable planets will orbit close to the star where they should transition from sub-Alfvenic to superAlfvenic regimes on every planetary orbit. Joule heating in the atmosphere of such a magnetized planet is significant (albeit lower than an unmagnetized planet), especially when the planet is an a sub-Alfvenic regime. At Earth, Joule heating is generally correlated with increased atmospheric ion escape. Dong et al. (2017) simulated an Earth-like planet orbiting orbiting Proxima Centauri-b, and found that the presence of a magnetic field generally lowers atmospheric ion escape rates. And Dong et al. (2020) simulated magnetized and unmagnetized Venus and Earth analogs orbiting TOI-700d. The magnetized Earth-like planets in their simulations had escape rates that were intermediate between the unmagnetized Earth analog and the unmagnetized Venus analog. Collectively, the studies above suggest that planetary magnetic fields influence escape rates for at least some planets, but not necessarily in obvious and consistent ways.

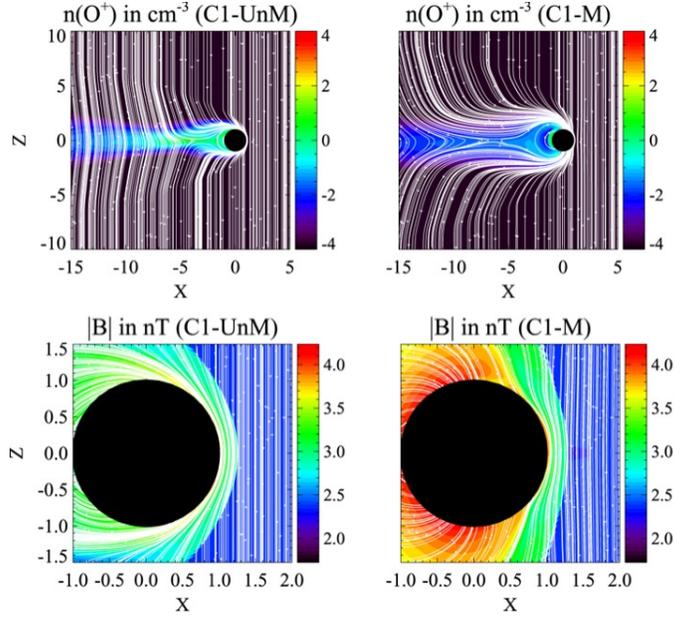

**Figure 18.** MHD simulation results for an unmagnetized (left panels) and magnetized (right panels) approximately Venus-like planet orbiting Proxima Centauri b. The assumed magnetic field strength for the results in the right panels is 1/3 Earth's. The star is to the left in all panels. The top panels show the atomic oxygen ion density, with white magnetic field lines. The bottom panels show close-ups of the magnetic field strength near the planet. From Dong et al. (2017).

**Particle Precipitation**

Several different populations of charged particles can precipitate into a planet's atmosphere or onto its surface. These include Galactic Cosmic Rays (GCRs), stellar wind particles, Stellar Energetic Particles (StEPs), and particles accelerated within the planet's magnetosphere. Each population of particles has a different range of energies, with typically different consequences for the planet. GCRs originate from a variety of sources outside of the astrosphere of the planet's host star, are highly energetic, and are commonly associated with supernovae and active galactic nuclei. The stellar wind originates in the stellar corona of the host star, contains both electrons and ions, and is less energetic than GCRs. StEPs also originate in the stellar corona as well as in the stellar wind, are more energetic than the stellar wind, and are associated with energetic events from the star such as coronal mass ejections and flares. To this point, many exoplanets have been identified around active stars, orbiting at small distances; this is likely to make the flux of StEPs large relative to the fluxes experienced at Earth (FFraschetti and RJolitz, submitted 2023). Finally, particles within a planet's magnetosphere (induced or intrinsic) can be accelerated by processes such as field-aligned currents, magnetic reconnection within the magnetosphere, or interactions with orbiting moons. Some of these particles may be accelerated into the planet's atmosphere, causing heating, ionization, or aurora. Collectively, precipitating particles can have energies ranging from $\sim 10^0$ eV up to $\sim 10^{20}$ eV or more.

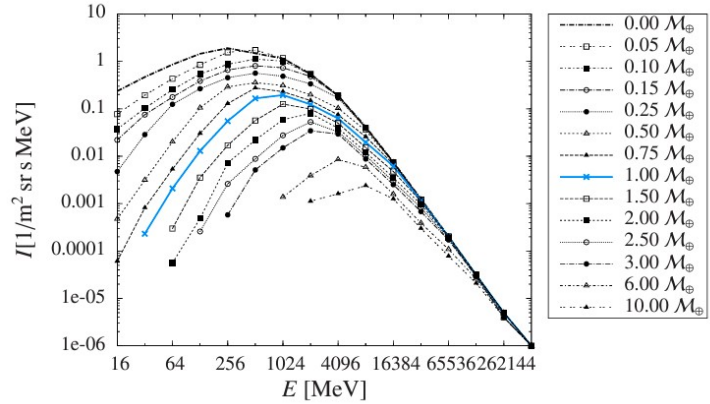

Figure 19: Flux of GCRs incident upon a planet orbiting a Sun-type star as a function of particle energy. Different curves correspond to different planetary dipole field strengths. From Grießmeier et al. (2015).

Magnetic fields alter the trajectory of a charged particle, with turning radius given by

$$r = \frac{m\, v_\perp}{q\, B}. \tag{14}$$

Thus, the incident energy spectrum of precipitating particles is modified by the presence of a magnetic field, with low energy and low mass particles more strongly attenuated than high energy particles. As an example, Figure 18 shows calculated energy spectra of GCR fluxes precipitating into the atmospheres of Earth-like planets with different magnetic field strengths. For sufficiently strong planetary fields, the GCR fluxes are reduced by a factor of 1000 or more. The detailed energy spectrum of precipitating GCRs, stellar wind particles, and StEPs depend upon many factors, but in general an exoplanetary magnetic field should reduce the precipitating fluxes from these sources. At the same time, locally accelerated particles may be more abundant and more energetic near magnetized planets, due to the stronger fields and currents associated with intrinsic magnetospheres.

Precipitating particles can have several consequences for planets, from their magnetosphere to their surface. Sufficiently strongly magnetized planets such as Jupiter (or even Earth) have radiation belts, comprised mostly of energetic charged particles captured from the stellar wind. If a planet has a magnetic field, then it may have a radiation belt. In some cases, this may be how an exoplanet's magnetic field is measured (see the Radiation Belt Emission subsection above).

Precipitating particles can also influence the chemistry of planetary atmospheres, affecting the abundance of climatically important trace gases. An example is the production of nitrogen oxides in Earth-like atmospheres by StEP protons with energies of several 10's of MeV (Segura et al., 2010). During stellar flares, production of nitrogen oxides should increase, as it does on Earth (Rodger et al., 2008). This is associated with a corresponding decrease in oxide molecules, including ozone (Segura et al., 2010). One may expect similar effects in the atmospheres of water worlds, since flares at Earth are also associated with the production of hydrogen oxides (Rodger et al., 2008).

The radiation reaching the surface of a rocky exoplanet (and therefore its habitability) can be significantly reduced by the presence of a planetary field due to the deflection of incident precipitating particles. This is not guaranteed, however. The amount of radiation reaching the surface depends upon both the strength of the field and the thickness of the planet's atmosphere. For a sufficiently thick atmosphere (or a sufficiently weak field), the surface radiation will not be significantly influenced by a planetary magnetic field. For example, Molina-Cuberos et al. (2001) calculated that the presence of an Earth-like magnetic field at Mars would not alter the radiation dose at the surface.

An indirect influence of precipitating particles on surface radiation may occur because of the chemistry discussed above. Increase production of nitrogen oxides, causing decreases in ozone abundance, would enable greater fluxes of damaging UVB radiation (Tilley et al., 2019) to reach a planet's surface, influencing habitability.

Lastly, rocky planets without significant atmospheres will experience surface sputtering in the absence of a sufficiently strong magnetic field. Surface sputtering occurs at both the Moon and Mercury today, as well as at asteroids and many outer planet satellites. The interaction between the surface and charged particles can create an observable exosphere as at the Moon or Mercury (Wurz et al., 2007; Lammer & Bauer, 1997), alter surface chemistry as at Europa (Ip et al., 1998), and change the color of the surface as at Ganymede (Ip et al., 1997). Ganymede presents an especially interesting case because it possesses a global magnetic field, and the surface sputtering occurs where charged particles are accelerated from Jupiter's magnetosphere along open magnetic field lines to the surface.

**Rotation Rate and Orbit**

An exoplanetary magnetic field can influence the motion of the planet itself. It is thought that giant planets should form with high rotation rates, near their 'breakup velocity', as they accumulate angular momentum from the material that accretes to form the planet. The rotation rates of gas giant exoplanets are likely to be reduced by the presence of a magnetic field, if the field formed while a partially ionized gas disk was still present near the planet on which the planet's magnetic field can exert torque, dissipating momentum into the disk. This magnetic braking effect was first proposed for Jupiter and Saturn, which rotate much more slowly than their breakup velocity (Takata & Stevenson, 1996). The idea was expanded upon by Batygin (2018), who asserted that the internal energy forming giant planets would both contribute to planetary magnetic field generation and provide luminosity that partially ionizes the disk, allowing efficient and rapid magnetic braking to occur. Observations of planetary mass objects support these ideas (Figure 19), with typical rotation rates well below the breakup velocity (Bryan et al., 2020). Ginzburg & Chiang (2020) showed that the timescale for magnetic braking is shorter than the timescale for giant planet contraction, so that giant planets never approach their breakup velocity. In fact, giant planets may increase their rotation somewhat after the surrounding gas disk dissipates and further contraction occurs.

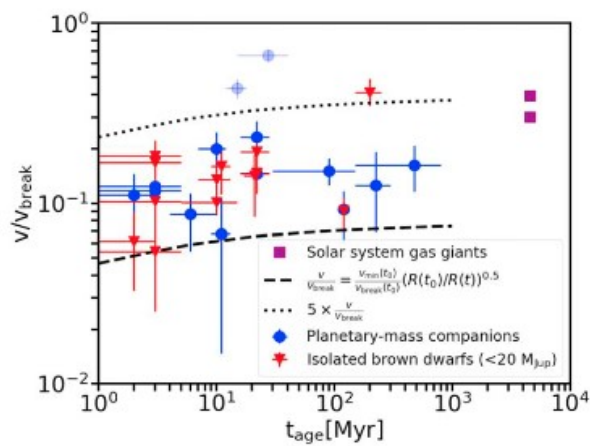

Figure 20: Rotation velocities relative to the break-up velocity as a function of age for a variety of objects, including Jupiter and Saturn, brown dwarfs, and planetary mass objects. Rotation rates are determined from spectral line broadening. From Bryan et al. (2020), reproduced by permission of the AAS.

Magnetic torques can be exerted between a close-in planet's magnetic field and its host star or stellar wind. When the wind is sub-Alfvénic, such planets can exert torques directly on the star. The resulting exchange of angular momentum can change the rotation of the star, causing the planet to migrate inward or outward (depending upon the distance of the planet from the star). In some cases, this effect is sufficiently strong that it dominates tidal effects between the planet and star (Strugarek et al., 2017). This situation is applicable to the kind of systems discussed in the Star-Planet Interactions subsection above: close-in planets magnetically connected to their star.

# ACKNOWLEDGMENTS

We thank J.S. Pineda, H. Beltz, E. Schreyer, and R. Murray-Clay for valuable feedback and discussions. The authors led sections as follows: D.A. Brain (Introduction and Influences sections); M.M. Kao (Assessing section); J.GO'Rourke (Theory section). D.A. Brain acknowledges support from NASA grants 80NSSC20K0594 and 80NSSC23K1358. M.M. Kao acknowledges support from the Heising-Simons Foundation through the 51 Pegasi b Fellowship grant 2021-2943.